\documentclass[review]{elsarticle}

\usepackage{lineno,hyperref}
\usepackage{color,soul}
\modulolinenumbers[5]

\journal{Journal of \LaTeX\ Templates}

\begin{document}

\begin{frontmatter}

\title{HaptStarter: Designing Haptic Stimulus Start System for Deaf and Hard of Hearing Sprinters}

\author[mainaddress]{Akihisa Shitara}
\address[mainaddress]{University of Tsukuba, 1-2 Kasuga, Tsukuba, Ibaraki, Japan}
\ead{Corresponding author}
\cortext[correspondingauthor]{Corresponding author}
\ead{theta-akihisa@digitalnature.slis.tsukuba.ac.jp}

\author[secondaryaddress]{Miki Namatame}
\address[secondaryaddress]{Kyoto Women's University, Faculty of Home Economics, 35 Kitahiyoshi-cho, Imakumano, Higashiyama-ku, Kyoto, Japan}

\author[thirdaryaddress]{Sayan Sarcar}
\address[thirdaryaddress]{School of Computing and Digital Technology, Birmingham City University, 15 Bartholomew Row, Birmingham, UK}

\author[fourthaddress]{Yoichi Ochiai}
\address[fourthaddress]{Research and Development Center for Digital Nature, University of Tsukuba, 1-2 Kasuga, Tsukuba, Ibaraki, Japan}

\author[fifthaddress]{Yuhki Shiraishi}
\address[fifthaddress]{Department of Industrial Information, Faculty of Industrial Technology, Tsukuba University of Technology,  4-3-15 Amakubo, Tsukuba, Ibaraki, Japan}

\begin{abstract}
 In this study, we design and develop HaptStarter---a haptic stimulus start system---to improve the starting performance of the deaf and hard of hearing (DHH) sprinters. A DHH person has a physical ability nearly equivalent to hearing; however, the difficulties in perceiving audio information lead to differences in their performance in sports. 
 Furthermore, the visual reaction time is slower than the auditory reaction time (ART), while the haptic reaction time is equivalent to it. 
 However, a light stimulus start system is increasingly being used in sprint races to aid DHH sprinters. In this study, we design a brand-new haptic stimulus start system for DHH sprinters; we also determine and leverage an optimum haptic stimulus interface. The proposed method has the potential to contribute toward the development of prototypes based on the universal design principle for everyone (DHH, blind and low-vision, and other disabled sprinters with wheelchairs or artificial arms or legs, etc.) by focusing on the overlapping area of sports and disability with human--computer interaction.

\end{abstract}

\begin{keyword}
\texttt{Accessibility, Sports,  Deaf and Hard of Hearing,  Haptic Stimulus, Start}
\MSC[2010] 00-01\sep  99-00
\end{keyword}

\end{frontmatter}


\section{Introduction}
 Globally, about 466 million people are deaf and hard of hearing (DHH); 34 million are children~\cite{Who:Deafness}. The communication and education support for the DHH people are facilitated mostly through hearing aids~\cite{Dillon:HearingAids}, cochlear implants~\cite{Clark:Cochlear}, automated speech recognition~\cite{Kheir:ASR}, and subtitles~\cite{Zoe:Subtitles}.
 However, in the sports domain, there are few support technologies for DHH people, which makes it difficult for them to actively participate in sports competitions such as the Olympics and Paralympics. They are forced to participate only in deafsports competitions such as the Deaflympics~\cite{ICSD:Deaflympics}. 
 The current scenarios are described below: DHH people participating in a sprint race, problems that occur based on the rules, and devices used to support the start of the race.

\subsection{Deaf Sports and Deaflympics}
 DHH people can participate alongside people without disabilities in most sports; however, it is difficult to rely only on audio information as the auditory stimuli, which are used as the standard for the start signal and referee’s signal in competitive sports. Therefore, assistive devices such as hearing aids and cochlear implants are often used to make it easier to perceive auditory cues for DHH people~\cite{Palmer:DeafAthlete}. Thus far, the physical abilities of DHH people have been studied rigorously~\cite{Stewart:Deaf, Myklebust:PerfomanceDHHChildren, Okuzumi:DHHBalance}. However, all of these studies report that DHH people face problems in communication compared to hearing people (e.g., coaches and teammates) because of their limited ability to perceive audio information~\cite{Saito:PerformanceDHH,Palmer:DeafAthlete}.
 
 Thus, in deaf sports, the cues, such as start and referee signals, etc., are leveraged as a visual stimulus instead of the conventional auditory stimulus for solving this problem. A representative example is the Deaflympics rule that specifies visual stimulus usage, which is quoted below from the text of DG23 ``AUTHORITY \& JURISDICTION(REFEREES, UMPIRES, \& JUDGES)'' - 3~\cite{ICSD:VisualRules}.
 
 \begin{quote}
  The competition rules for each sport shall be those of the International Federations as amended where visual cues are to be used in place of auditory cues. 
 \end{quote}
 
 Another case study reported used the visual stimulus at the Deaf Athlete Game in Malaysia~\cite{xZulkiflli:MalaysiaVisual}.

\subsection{Visual Stimulus of the Physical Characteristics Problem}
 DHH sprinters, while participating in a race with the traditional system that uses the pistol sound as the start signal, face the following difficulties:
 
 \begin{enumerate}
    \item listening to the start sound signal can cause anxiety
    \item continuously looking at the starter to perceive the visual cue
    \item simultaneously looking at the starter and the sprinter's movement in the next lane
 \end{enumerate}
 
\begin{figure}
  \centering
  \includegraphics[width=1.0\columnwidth]{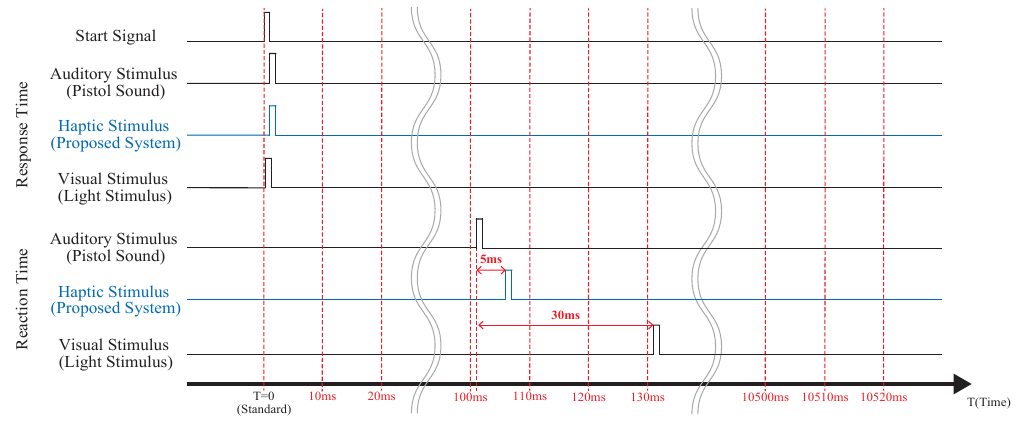}
  \caption{Relationship between the minimum unit of measurement of the sprint race and Ifukube’s report about reaction time delay based on auditory reaction time.}~\label{fig:IufukubeReport}
 \end{figure}

 The light stimulus start system~\cite{NISHI:OpticalStart} is an example of a system that can solve such problems; DHH athletes currently use this system when participating in sprint races.
 However, prior research showed that the visual reaction time (VRT) of humans is slower than their auditory reaction time (ART)~\cite{Welford:ReactionTimes,Woodworth:ExperimentalPsyhology}. Furthermore, VRT is slower than the ART by approximately 30 ms~\cite{Ifukube:SensoryProsthesis} (Figure~\ref{fig:IufukubeReport}). 
 Another study on RT measurement for basketball players reported that VRT is slower than ART~\cite{Ghuntla:Basketball-AudioVisualRT}.
 
 As an example, consider that DHH sprinters take the same time as able-bodied sprinters to run a 100 m sprint race. However, the DHH sprinters use a light stimulus start system to start the race, while the able-bodied sprinters follow the pistol sound as the start signal. Consequently, the reaction time of the DHH sprinter becomes slower (average of 30 ms) than the able-bodied sprinters.
 This start timing delay time affects the sprint race's final times~\cite{Ana:RTandSpritRace,Aditi:PerformanceSprint}. Among others, the men's 100 m sprint world record time is 9'' 58 s~\cite{IAAF:100m}. However, the DHH men's 100 m sprint world record time is not far, and is only 10'' 21 s~\cite{ICSD:100m}.
 In addition, the start of a DHH sprinter can also be delayed because of missing the light stimulus due to an ``eyeblink'' because each eyeblink is about 100 ms of blackout time~\cite{Volkmann:eyeblink}. The relationship between blinking and sports performance has been reported~\cite{Yoshida:eyeblink,Ishigaki:eyeblink}. Therefore, if the light of the start signal is triggered while the DHH is blinking, the DHH may misses the start timing and lag in the race because the blackout time exceeds the minimum measurement unit of 10 ms. 
 Moreover, such assistive devices are less likely to be promoted in standard sports~\cite{UN:Disability} as inclusive technology accessible for every sportsperson.
 
 In this study, we design and develop an alternate stimulus-based sprint race start system suitable for DHH sprinters. Prior research studies show that the VRT of humans is slower than the ART, and haptic RT is faster than VRT and nearly equivalent to the ART~\cite{Welford:ReactionTimes,Woodworth:ExperimentalPsyhology}. Ifukube's report also stated that the haptic RT is slower than ART by approximately 5 ms~\cite{Ifukube:SensoryProsthesis} (Figure~\ref{fig:IufukubeReport}). Another study conducted at a university in Uganda reported that haptic RT is faster than ART and VRT. 
 Existing literature~\cite{Petermeijer:DriverRT, Gray:HapticRT_Drive, Itoh:HapticRT_Drive, Jordan1:HapticRT_Drive, Jordan2:HapticRT_Drive, Mohebbi:HapticRT_Drive, Straughn:HapticRT_Drive, Scott:HapticRT_Drive} also indicated that an RT is fast when using a haptic support system.
 Hokari and Miyamoto et al. developed a system that leverages haptic stimulus for DHH sportspersons~\cite{Hokari:WarningforSportsforDHH}. The system uses a vibration-type haptic stimulus to send notifications or warning sounds through a whistle. However, this system does not leverage the combined expertise in individual sports rules and scenes. Furthermore, such a system with a time lag is unsuitable for DHH participants in sports where the time of completion is the main factor for victory. For example, the sprint race time is recorded from the time of the start signal sound because the sprint race does not change the start signal, even if the DHH sprinter participates in a general race.
 Given this context, we explored haptic modality to improve the perception of the start signal in the sprint race for DHH sprinters. Furthermore, we aim to change the start signal to a single stimulus as a ``universal design'' principle for everyone, and not as an assistive technology tool for only DHH people.
 Therefore, we conducted an experiment targeting for DHH male people and also organized and considered our experiment's results and the background surrounding DHH to aim for our future goals.

\section{HaptStarter Prototype Design}
 In this section, we present the design of the \textit{HaptStarter} prototype based on two major parts: the device to generate haptic stimulus and the contact interface.
  
 \begin{figure}[t]
    \centering
    \includegraphics[width=1.0\columnwidth]{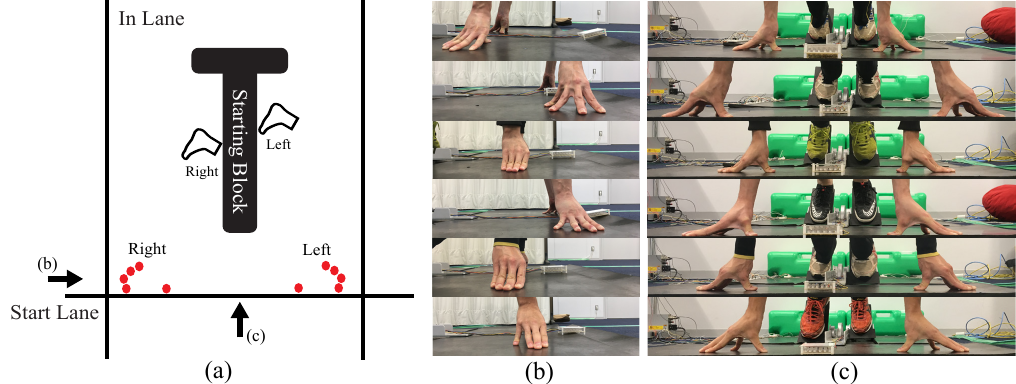}
    \caption{ (a) Posture of the sprinter when he is in the Set. (b) Side view of the sprinter. (c) Front view of the sprinter.}
    \label{fig:SetHand}
 \end{figure} 

\subsection{Haptic Stimulus Generating Device}
 We proposed and developed a start system that presents a haptic stimulus from direct signals suitable for DHH sprinters; the stimulus is specifically adapted to a crouch start. 
 In the method proposed in our first study~\cite{Shitara:VibStart}, the thumb and fingers of the sprinter touched the plates attached to vibration motors; the workflow involved vibrating the contacting plate to indicate the start signal.
 
 However, statistical analysis revealed no significant differences between the RT on light stimulus and vibration stimulus. However, the questionnaire and interview results presented two crucial points: ``The vibration device limits the thumb and finger placement'' and ``The strength of vibration is weak.'' Although the system could not solve the existing issues, the fastest time of the RT was achieved with the vibration stimulus, which indicated the potential of our first work.
 
 In our second study~\cite{Shitara:TacStart}, two haptic stimuli (push-type and vibration-type) were examined to identify the optimum haptic stimulus for DHH sprinters; the results confirmed that the push-type haptic stimulus was more suitable than the vibration-type. 
 The vibration-type and push-type devices include a motor and a plastic plate fabricated using a three-dimensional (3-D) printer. In the former, the motor vibrates the contacting plate, which is in contact with the thumbs the sprinter set. The contacting plate vibrates to the start signal. In the latter, the solenoid pushes the contacting plate, which is in contact with the thumbs the sprinter set. This workflow involves pushing the contacting plates to the start signal.
 
 A significant trend was observed between the mean values of the RT for the push-type and vibration-type devices. Furthermore, a significant difference was also observed between the standard deviation of the RT values for the push-type and vibration-type devices; this result implied that the push-type device was more stable than the vibration-type device.
 Although there was no significant difference between the mean of the RT for the push-type and LED-type devices, a significant trend was observed between the standard deviation of the RT for the push-type and LED-type devices. The standard deviation of the push-type stimuli was smaller than that of the LED-type one; thus, the push-type device was more stable than the LED-type, which implies that the push-type device is more suitable as the start signal for DHH sprinters than the LED-type. In addition, the questionnaire indicated that the push-type haptic stimulus was easier to recognize than the existing LED-type visual stimulus. The result of ``Which stimulus makes it easier to start?'' shows the limitation of the existing push-type device. Almost all participant responses indicated that the push power was still weak, and the restricted hand position was uncomfortable, which is similar to the result obtained in our first study. However, the responses to ``Which device has a higher future potential?'' indicated that all participants expect the future potential of the push-type to be higher than that of the LED-type device. 

 These research results focus on the hand placement of the athlete in sprint races. We focus on hand placement by elimination based on three points that include easy contact with a device, not wearing a device after starting, and ``RULE 162 The Start''~\cite{IAAF:RulesStart}, which is 

 \begin{quote}
    An athlete shall not touch either the start line or the ground in front of it with his hands or his feet when on his mark. Both hands and at least one knee shall be in contact with the ground and both feet in contact with the foot plates of the starting blocks. At the "Set" command, an athlete shall immediately rise to his final starting position retaining the contact of the hands with the ground and of the feet with the foot plates of the blocks.
  \end{quote}
  
 Before starting the experiment, the hand position, fingers, etc., vary in the usual start (when using LEDs), as shown in Figure~\ref{fig:SetHand}.
 The thumb is in a relatively stable position to maintain the center of gravity, and therefore, we employ the method that considers contact with the thumb.

\subsection{Contact Interface}
 Our first study reported the development of a ``vibration-type'' haptic device~\cite{Shitara:VibStart}. However, this device has a small advantage over other devices because it offers no significant difference in the faster reaction compared to that with the visual stimulus. Therefore, in our second study, we used the ``push-type'' and ``vibration-type'' from haptic stimulus types (e.g., ``electrical-type'' and ``air-type''); the results confirmed that the ``push-type'' is more suitable as the haptic stimulus transmission method for DHH sprinters~\cite{Shitara:TacStart}.  However, the ``push-type'' haptic device achieves no significant difference similar to the ``vibration-type'' haptic device used in our first study. 
 Furthermore, the device has a problem in that it is ``difficult to clearly receive the start signal'' because the contact interface of the second device presents a stimulus to the thumb using a method that presses a small object with a curved surface in abutting contact with the thumb.
 To solve these problems mentioned above, we need to consider the following points and improve a interface part based on the findings of previous ``push-type'' haptic device~\cite{Shitara:TacStart}:

 \begin{itemize}
    \item Easy to clearly perceive the start signal
    \item Easy to decide where someone sets a device
 \end{itemize}
 
 Therefore, we aimed to achieve the fastest RT using a strong stimulus to clearly perceive the start signal~\cite{Harry:Stimulus}. Thus, the distance to stroke the core of the solenoid was redesigned to 3 mm because the haptic device uses the solenoid (Shidengen Mechtaronics's small push-pull version 14C) used in the previous device~\cite{Tanaka:Solenoid}.

 \begin{figure}
    \centering
    \includegraphics[width=1.0\columnwidth]{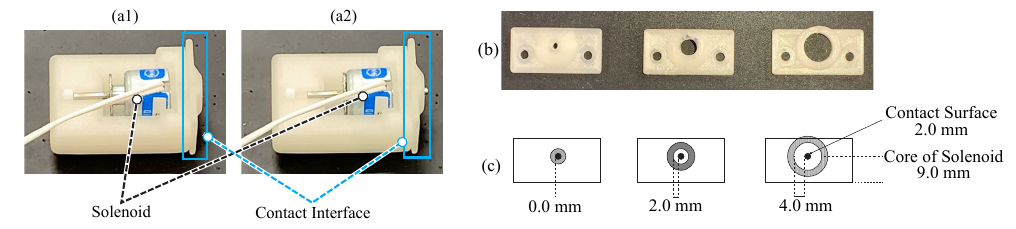}
    \caption{HaptStarter prototype that comprises a solenoid and contact interface. Behavior of haptic devices ((a1): Before start, (a2): After start). (b) Actual device and (c) schematic design of the three types of contact interfaces marked by the light blue rectangle in (a1) and (a2), respectively.}
    \label{fig:NewHaptic}
 \end{figure} 

 We also need to improve the previous contact interface that presents the stimulus to the thumb considering the two requirements: 1) easy to decide where someone sets a device and 2) easy to clearly perceive the start signal after someone sets a device. To this end, we redesigned the contact interface to a two-stage type from a one-stage type because the solenoid can touch the thumb of the sprinter even if it is separated slightly after sprinters set their thumbs in contact with the plate. However, the two-stage contact interface needs to allow the sprinter to clearly perceive the start signal. Therefore, to clearly perceive the difference between the ``set'' and ``start'' states when the device contacts a hand, we created three types of contact interfaces considering the distance between the contact surfaces and the  solenoid's core (0 mm, 2 mm, and 4 mm) (Figure~\ref{fig:NewHaptic}) using a 3D printer.
 Because we assumed that ability to perceive the difference between the “set” and “start” states when the device contacts a hand is the two-point discrimination acuity in the fingers.
 We considered the two-point discrimination acuity in the fingers to be 1 mm, as reported by Kimberly et al.~\cite{Kimberly:TactileModality}. Also, we decided on contact interfaces considering the distance between the contact surfaces and the solenoid's core to check at 2 mm intervals based on the Psychophysical Tactile Interaction Design Guidelines~\cite{Hale:HapticDesignGuidelines}.

\section{Experiment}
 Previous studies focused on simple RTs, such as during button pressing~\cite{Shelton:AudioVisualSimpleRT}. However, unlike previous research studies, the current study considers the whole-body RTs, such as when using a crouching start. Previous studies have measured and compared whole-body RTs to auditory and visual times; other studies have examined the relationship with simple RTs. To the best of our knowledge, no studies measure the body RT for haptic and auditory or haptic and visual stimuli. This study defines simple RTs as RTs by pressing a button, and whole-body RTs as RTs using a crouch start.

 Previous studies have shown that the finger pad of the thumb is impressionable~\cite{Kimberly:TactileModality}; however, it cannot be used because the finger pad of the thumb at the start of crouching is in contact with the ground. We adopted the contact method to the outside of the first joint to make it easier to move forward when starting. However, we cannot find any information on the first joint of the haptic sensation of the thumb. 

 Therefore, we assumed that the haptic interface cannot be used in the sprint race using the crouch start if the RT difference between the two contact points (outside the first joint of the thumb and the finger pad of the thumb), each with the same haptic interface, is significant. Therefore, we examined the RT pressing the button for the three-type haptic interface and LED-type visual stimulus in Study 1 to select the haptic interface, which is the smallest RT difference between the two contact points.
 
 However, when using a crouching start, the reaction time does not have the same result as when using a button; however, they are related. Therefore, we examined the RT using the crouch start for the optimum haptic interface and LED-type visual stimulus in Study 2 to determine if the optimum haptic stimulus interface for DHH sprinters is more effective than an LED-type stimulus in the crouch start.

 In addition, the previous studies~\cite{Wells:RT_Reliability_Performance,Del:RT_Practice} show that simple RTs are easily affected by practice effects and require three trials. Therefore, we decided that ten trials will be conducted per one time in the simple RT measurement experiment, and five trials will be conducted per stimulus for one time in the whole body RT measurement experiment.

 The research ethics committee of the Tsukuba University of Technology approved this experiment design.

\subsection{Participants}

\begin{table}[t]
    \caption{Participants in the experiment}
    \begin{tabular}{lccccc}
            \begin{tabular}{l}
                ID 
            \end{tabular} &
            \begin{tabular}{c}
                Age 
            \end{tabular} &
            \begin{tabular}{c}
                dB 
            \end{tabular} &
            \begin{tabular}{c}
                Athletics \\ 
                history \\ 
                (Years) 
            \end{tabular} &
            \begin{tabular}{c}
                Event
            \end{tabular} &
            \begin{tabular}{c}
                Personal best
            \end{tabular}
        \\
            \begin{tabular}{l}
                P1 
            \end{tabular} &
            \begin{tabular}{c}
                21 
            \end{tabular} &
            \begin{tabular}{c}
                R: 93.5 \\
                L: 106.3
            \end{tabular} &
            \begin{tabular}{c}
                3
            \end{tabular} &
            \begin{tabular}{c}
                200 m \\
                400 m
            \end{tabular} &
            \begin{tabular}{c}
                24" 98 \\
                54" 86
            \end{tabular}
        \\
        \hline
            \begin{tabular}{l}
                P2 
            \end{tabular} &
            \begin{tabular}{c}
                22 
            \end{tabular} &
            \begin{tabular}{c}
                R: 106.25 \\
                L: 110
            \end{tabular} &
            \begin{tabular}{c}
                8
            \end{tabular} &
            \begin{tabular}{c}
                100 m \\
                400 m
            \end{tabular} &
            \begin{tabular}{c}
                13" 00 \\
                62" 00 (units)
            \end{tabular}
        \\
        \hline
            \begin{tabular}{l}
                P3 
            \end{tabular} &
            \begin{tabular}{c}
                21 
            \end{tabular} &
            \begin{tabular}{c}
                R: 108 \\
                L: 108
            \end{tabular} &
            \begin{tabular}{c}
                8
            \end{tabular} &
            \begin{tabular}{c}
                100 m  \\ 
                Triple jump
            \end{tabular} &
            \begin{tabular}{c}
                12" 01  \\ 
                12 m 50 cm (units)
            \end{tabular}
        \\
        \hline
            \begin{tabular}{l}
                P4 
            \end{tabular} &
            \begin{tabular}{c}
                21 
            \end{tabular} &
            \begin{tabular}{c}
                R: 90--100 \\
                L: 90--100
            \end{tabular} &
            \begin{tabular}{c}
                6
            \end{tabular} &
            \begin{tabular}{c}
                100 m \\
                200 m
            \end{tabular} &
            \begin{tabular}{c}
                11" 87 \\
                23" 95
            \end{tabular}
        \\
        \hline
            \begin{tabular}{l}
                P5 
            \end{tabular} &
            \begin{tabular}{c}
                22 
            \end{tabular} &
            \begin{tabular}{c}
                R: 80 \\
                L: 90
            \end{tabular} &
            \begin{tabular}{c}
                3
            \end{tabular} &
            \begin{tabular}{c}
                100 m  \\ 
                Long jump
            \end{tabular} &
            \begin{tabular}{c}
                12" 3  \\ 
                5 m 70 cm (units)
            \end{tabular}
        \\
        \hline
            \begin{tabular}{l}
                P6 
            \end{tabular} &
            \begin{tabular}{c}
                22 
            \end{tabular} &
            \begin{tabular}{c}
                R: 92 \\
                L: 103
            \end{tabular} &
            \begin{tabular}{c}
                8
            \end{tabular} &
            \begin{tabular}{c}
                100 m \\
                200 m \\ 
                400 m
            \end{tabular} &
            \begin{tabular}{c}
                12" 10 \\
                23" 00 (units) \\ 
                53" 00 (units)
            \end{tabular}
        \\
    \end{tabular}
    \label{tb:participants}
\end{table}
 
 We recruited six DHH participants (all males) with more than three years of the athletics sprint experience at Tsukuba University of Technology; the details of all participants are indicated in Table~\ref{tb:participants}. The participants are, on average, 21.5 years old (SD = 0.5, range = 21--22). The participants reported hearing loss as congenital; furthermore, we not only needed to recruit DHH people who had experience and were accustomed to crouch start in the sprint race of athletics, we also required to make the same participants of Studies 1 and 2. We considered the following three factors: the significant difference caused by the habituation of crouching start~\cite{Yau:Visual_Athletes_NonAthletes} and the habituation to the present stimulus~\cite{Christopher:SwimmingStartRT}. We limited ourselves to male participants because 1) it is difficult to collect the same number of females who have the experience and are accustomed to crouch start in the sprint race, and 2) the suppression of the significant difference caused by gender~\cite{Ferguson:RT_SexBody}. We also asked about other details such as athletics history, athletic events, and personal best of athletic events because our study focuses on athletics.

\subsection{Study 1: Reaction Time Using the Button Press}
 We assumed no significance between the finger pad and the first joint of the thumb to identify the optimum haptic stimulus interface for DHH sprinters, also examined the RT using the button press for a combination of six (contact interface: three types, contact point: two positions (Figure~\ref{fig:WayHolding}) types of haptic stimulus interface.

\subsubsection{Apparatus}

    \begin{figure}[ht]
        \centering
        \includegraphics[width=\linewidth]{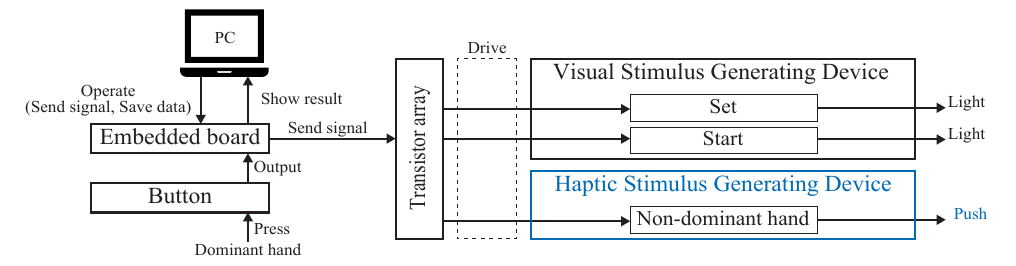}
        \caption{Measurement system using the button press}
        \label{fig:Method-ButtonPress}
    \end{figure} 

    \begin{figure}
        \centering
        \includegraphics[width=0.5\columnwidth]{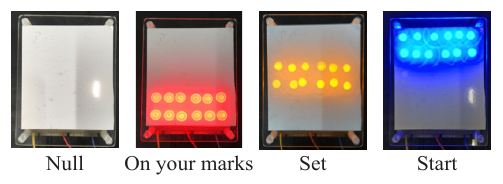}
        \caption{LED-type visual stimulus generating device}~\label{fig:LED-Device}
    \end{figure}

 Figure~\ref{fig:Method-ButtonPress} shows the state operation of the system using a button module. The RT measurement algorithm measures the time from when a start signal is emitted until the button is pressed for each the minimum time 1 ms. The start signal is emitted between 2000 ms and 3000 ms after operating. Furthermore, we examined the RT of the LED-type visual stimulus and compared it with the optimum haptic stimulus interface to confirm the effectiveness of the haptic stimulus on simple reaction times. Moreover, the LED-type visual stimulus generating device (Figure~\ref{fig:LED-Device}) is ``Yellow'' with ``Set'' from pressing one button to pressing one button. 

\subsubsection{Procedure}
 We conducted ten trial measurements for each of the six types of combination haptic stimulus interfaces and the LED-type visual stimulus per participant in one time. 
 The procedure for one trial measurement is as follows. The participants sitting in a chair and looking at the LED-type visual stimulus ``Yellow'' are stimulated with a haptic stimulus to the non-dominant hand or the visual stimulus ``Blue''; furthermore, the participants pressed the button with the dominant hand when recognizing the presented stimulus. 
 We repeated four times the above contents per participant.
 The participants experienced seven stimuli in random order and counterbalanced across participants and procedure times; the order of contact points was alternated each time a haptic stimulus was presented.
 
    \begin{figure}[ht]
        \centering
        \includegraphics[width=0.5\columnwidth]{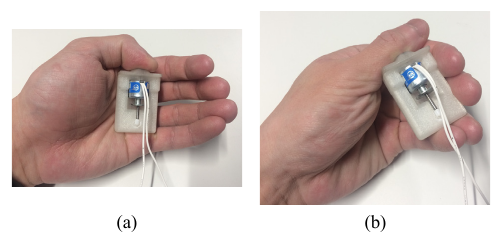}
        \caption{Approach to holding the haptic device}~\label{fig:WayHolding}
    \end{figure}
 
    \begin{figure}[ht]
        \centering
        \includegraphics[width=\linewidth]{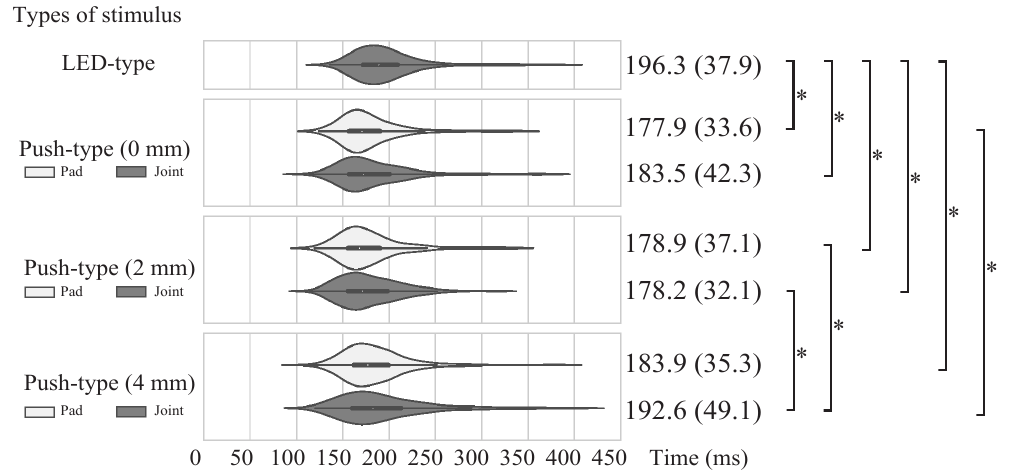}
        \caption{Violin plots of RT using the button press for seven experimental conditions. Mean and standard of RT using button press (``*'' is a comparison while a significant difference shown by both Tukey and Bonferroni tests)}~\label{fig:ResultButton}
    \end{figure}
 
 In addition, to examine no significant between the finger pad and the first joint of the thumb, in the case of contact with the finger pad of the thumb, Figure~\ref{fig:WayHolding} (a) shows how to hold the haptic stimulus generating device. Similarly, Figure~\ref{fig:WayHolding} (b) shows how to hold the device in the case of contact with the first joint of the thumb.

\subsubsection{Result}
 In this experiment, we collected 1680 data as the total data; 40 data were obtained from one stimulus per participant.
 The mean and standard deviation of the RT of the six combination types of haptic stimulus interface and LED-type visual stimulus are presented in Figure~\ref{fig:ResultButton}. 
 We conducted Shapiro-Wilk on the seven types of combinations; the distribution departed significantly from normality ($W = 0.86, p < 0.05$). 
 Therefore, we conducted two multiple comparison test that do not need to use ANOVA: Tukey-Kramer and Bonferroni. The results show ``*'' as a comparison representing a significant difference ($p < .001$) with both Tukey-Kramer and Bonferroni plots in the graph of Figure~\ref{fig:ResultButton}. The results indicate that the method involving a 0.0 mm push-type contact with the finger pad of the thumb is the fastest. However, the finger pad cannot be used in the crouch start in sprint races because it is in contact with the ground. The second fastest is the 2.0 mm push-type contact with the first joint of the thumb. 
 There is no significant difference between the method that uses a 2.0 mm push-type contact with the finger pad of the thumb. The 4.0 mm push-type contact with the first joint of the thumb cannot be used because this method shows a significant difference compared to the method that uses 2.0 mm and 0.0 mm push-type contacts with the finger pad of the thumb. Based on the results of this analysis, we adopt the method that uses a 2.0 mm push-type contact with the first joint of the thumb. 

\begin{table}[h]
    \begin{tabular}{clcc}
            \begin{tabular}{c}
                Order
            \end{tabular} &
            \begin{tabular}{l}
                Push-type
            \end{tabular} &
            \begin{tabular}{c}
                Median
            \end{tabular} &
            \begin{tabular}{c}
                Mean \\
                (Standard deviation)
            \end{tabular} 
        \\
            \begin{tabular}{c}
                1st
            \end{tabular} &
            \begin{tabular}{l}
                2.0 mm - Finger pad 
            \end{tabular} &
            \begin{tabular}{c}
                4.5
            \end{tabular} &
            \begin{tabular}{c}
                4.0 (1.7)
            \end{tabular}
        \\
        \hline
            \begin{tabular}{c}
                2nd
            \end{tabular} &
            \begin{tabular}{l}
                0.0 mm - Finger pad 
            \end{tabular} &
            \begin{tabular}{c}
                4.0
            \end{tabular} &
            \begin{tabular}{c}
                3.7 (1.7)
            \end{tabular} 
        \\
        \hline
            \begin{tabular}{c}
                3rd
            \end{tabular} &
            \begin{tabular}{l}
                4.0 mm - Finger pad 
            \end{tabular} &
            \begin{tabular}{c}
                4.0
            \end{tabular} &
            \begin{tabular}{c}
                3.7 (1.8)
            \end{tabular}
        \\
        \hline
            \begin{tabular}{c}
                4th
            \end{tabular} &
            \begin{tabular}{l}
                2.0 mm - First joint 
            \end{tabular} &
            \begin{tabular}{c}
                3.5
            \end{tabular} &
            \begin{tabular}{c}
                4.0 (1.5)
            \end{tabular} 
        \\
        \hline
            \begin{tabular}{c}
                5th
            \end{tabular} &
            \begin{tabular}{l}
                4.0 mm - First joint 
            \end{tabular} &
            \begin{tabular}{c}
                3.0
            \end{tabular} &
            \begin{tabular}{c}
                3.0 (1.6)
            \end{tabular} 
        \\
        \hline
            \begin{tabular}{c}
                6th
            \end{tabular} &
            \begin{tabular}{l}
                0.0 mm - First joint 
            \end{tabular} &
            \begin{tabular}{c}
                2.5
            \end{tabular} &
            \begin{tabular}{c}
                2.7 (1.5)
            \end{tabular}
        \\
    \end{tabular}
    \caption{The result of in order to ``Ease of recognize'' by the participants (1st: 6point -- 6th: 1point). Mean and standard deviation of the result shows calculated evaluation points by participants.}~\label{tb:Button-7_Liker-scale}
\end{table}

 Furthermore, the results of the questionnaire are presented in Table~\ref{tb:Button-7_Liker-scale}. The result indicates that the content is arranged in the order of the ``Ease of recognition'' for the six combinations of the haptic stimuli by the participants (1st: 6 point – 6th: 1 point). As indicated in the result, the first, second, and third highest are both the method contact with the thumb’s ``Finger pad.’’ The 2.0 mm push-type device is the highest for both ``Finger pad'' and ``First joint,'' and the 2.0 mm is the smallest in the difference of the ``Finger pad'' and ``First joint.'' Almost all participant responses mentioned that 2.0 mm would be better than the 0.0 mm of the push-type. 
 Therefore, the questionnaire result is close to the result of RT using the button press.

\subsection{Study 2: Reaction Time Using the Crouch Start}
 We examined the RT using the crouch start for the optimum haptic interface and LED-type visual stimulus to determine if the optimum haptic stimulus interface for the DHH sprinters is more effective than an LED-type stimulus.

\subsubsection{Apparatus}

    \begin{figure}[ht]
        \centering
        \includegraphics[width=\linewidth]{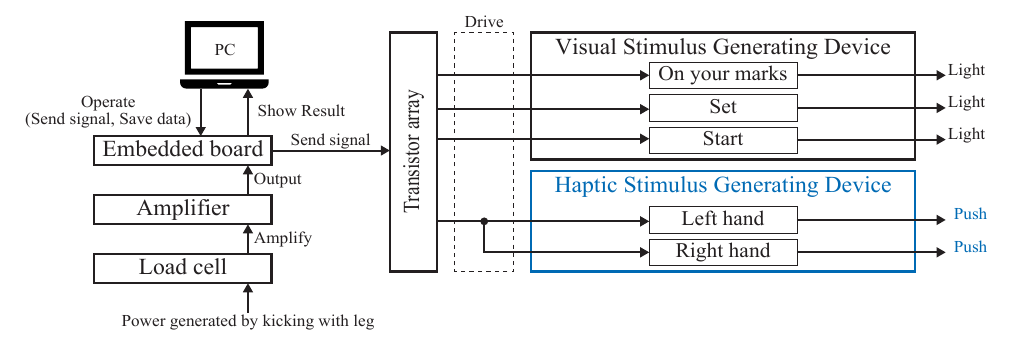}
        \caption{Measurement system using the crouch start}
        \label{fig:Method-CrouchStart}
    \end{figure} 

 Figure~\ref{fig:Method-CrouchStart} shows the state operation of the system using Yokokura's algorithm~\cite{Yokokura:StartAction} with a load cell placed under the starting block that receives the force kicking power. The RT measurement algorithm measures the time from when a start signal is emitted until the foot kicking force is transmitted to the load cell for each the minimum time 1 ms. 

    \begin{figure}[ht]
        \centering
        \includegraphics[width=\linewidth]{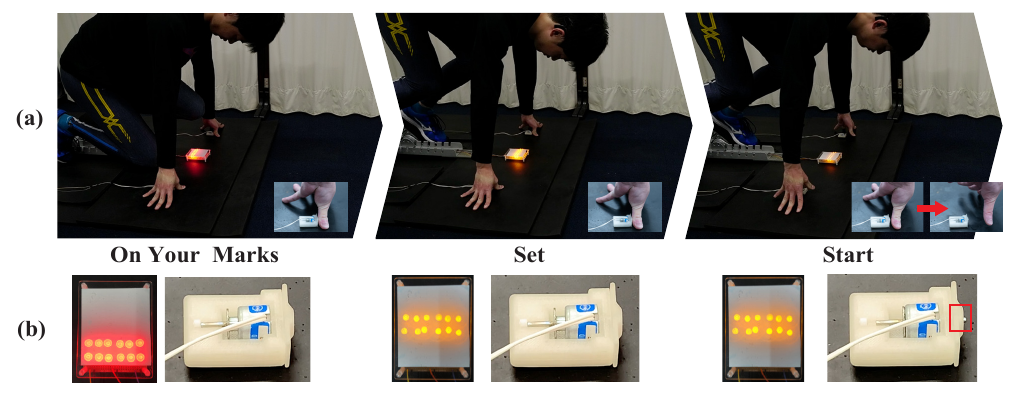}
        \caption{(a) Operation of HaptStarter and sprinter. (a) Overall views; the insets show the contact between the thumb and the HaptStarter. (b) Operation of the HaptStarter; the red square indicates the iron core of the solenoid pushing and contacting the thumb.}
        \label{fig:Prosee-CrouchStart}
    \end{figure}  

 The RT measurements are conducted based on the rule of start in sprint races~\cite{IAAF:RulesStart}; i.e., we indicate the movement in the order of ``On your marks,'' ``Set,'' and ``Start'' (each then, the stimulus occurs) and measure the RT in ``Start'' during the trial. The trial ended after the participant finished the start operation and ran about 5 m. The haptic stimulus generating devices only releases in the case of ``Start,'' the ``Red'' and ``Yellow'' LED-type visual stimulus generating device (Figure~\ref{fig:LED-Device}) corresponds to ``On your marks'' and ``Set,'' respectively Figure~\ref{fig:Prosee-CrouchStart}.
 
    \begin{figure}[ht]
        \centering
        \includegraphics[width=0.5\columnwidth]{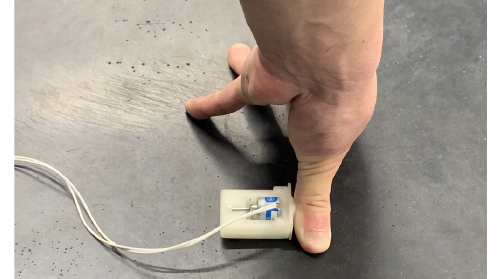}
        \caption{Position of the hand to be in contact with the haptic device.}
        \label{fig:WaySet}
    \end{figure} 

 Furthermore, before starting the RT measurement, we asked the participant to set the position where it is best to contact the first joint of the thumb in the movement ``On your marks'' or ``Set'' in the scene using the crouch start (Figure~\ref{fig:WaySet}).

\subsubsection{Procedure}
 We examined the RT using the crouch start of the employed contact interface on the push-type haptic stimulus and LED-type visual stimulus per participant.
 In this experiment, we performed five trial measurements for each of the two stimuli, and we measured the RT again if the participant did not satisfy the crouch start by self-reporting. 
 We repeated the above procedure 16 times, determined from the factorial of stimulus numbers.
 The participants experienced two stimuli in random order to counterbalance the results across participants and procedure times.
 We investigated the ``Ease of Start'' and ``Ease of Recognition'' with a seven-point Likert scale (1: worst, 7: best) via a questionnaire and interview after finishing the 8th and 16th procedure time of the RT using the crouch start.

\subsubsection{Result}

    \begin{figure}[ht]
        \centering
        \includegraphics[width=\linewidth]{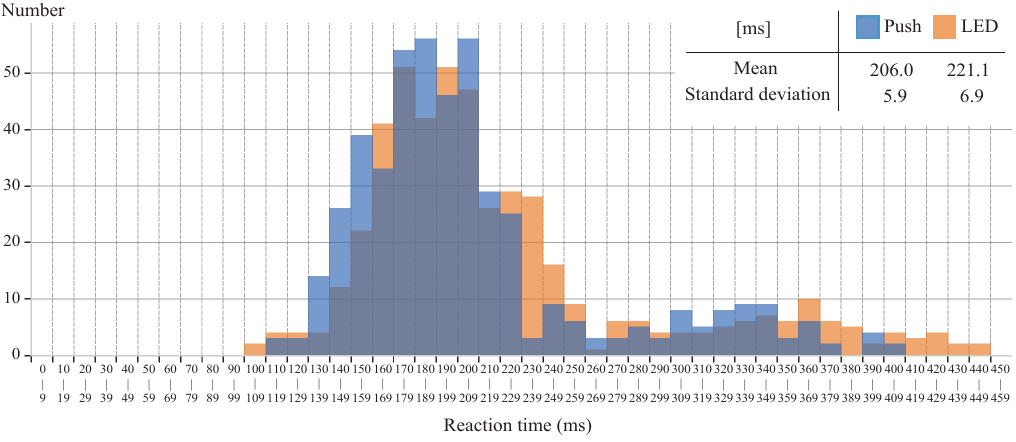}
        \caption{Histogram of reaction time using the crouch start for two stimuli. The mean and standard deviation of the RT using the crouch start are presented}~\label{fig:ResultStart}
    \end{figure}

 Five data were obtained through one procedure time for each of the two stimuli in one experiment per DHH participant. Therefore, the total number of data was 960 (LED: 480, Push: 480); 480 data is multiplied per stimulus by the number of six participants, five trial measurements, and procedure 16 times. The outliers of the obtained data were detected using standard deviation. We excluded data that deviated from three times the standard deviation range to the mean of the respective stimulus for each participant. We excluded 13 data (Push: 8 LED: 5). Figure~\ref{fig:ResultStart} shows the mean and standard deviation of each RT of the two stimuli.
 The results of Welch's t-test ($p = .0003 < .05$) indicated a significant difference between the mean of the RT of the push-type and that of LED-type. 
 Furthermore, the results of the F-test ($p = .0003 < .05$) indicated a significant difference between the standard deviation of the RT of the push-type and that of the LED-type, and we also confirmed there was a significant trend between the mean of the RT of the push-type and that of the LED-type in each participant (five person: $p < .05$, one person: $p < .10$).
 Thus, the push-type device was more stable than the LED-type; the push-type haptic stimulus device befits the start signal for the DHH sprinters better than the LED-type. 

\begin{table}[ht]
    \begin{tabular}{ccccc}
            \begin{tabular}{c}
            \end{tabular} &
            \begin{tabular}{c}
            \end{tabular} &
            \begin{tabular}{c}
                Median
            \end{tabular} &
            \begin{tabular}{c}
                Mean (SD)
            \end{tabular} 
        \\
            \begin{tabular}{c}
                ``Ease of recognition'' in LED\\
                (1: very bad - 7: very good)
            \end{tabular} &
            \begin{tabular}{c}
                 8th \\
                 16th 
            \end{tabular} &
            \begin{tabular}{c}
                6.0 \\
                6.0 
            \end{tabular} &
            \begin{tabular}{c}
                5.5 (1.3)\\
                5.8 (0.7)
            \end{tabular} 
        \\
        \hline
            \begin{tabular}{c}
                ``Ease of recognition'' in PUSH\\
                (1: very bad - 7: very good)
            \end{tabular} &
            \begin{tabular}{c}
                 8th \\
                 16th 
            \end{tabular} &
            \begin{tabular}{c}
                5.5 \\
                6.0 
            \end{tabular} &
            \begin{tabular}{c}
                5.2 (1.1)\\
                6.0 (0.8)
            \end{tabular} 
        \\
        \hline
            \begin{tabular}{c}
                ``Which is easier to recognize?''\\
                (1: LED - 7: PUSH)
            \end{tabular} &
            \begin{tabular}{c}
                 8th \\
                 16th 
            \end{tabular} &
            \begin{tabular}{c}
                4.5 \\
                4.5 
            \end{tabular} &
            \begin{tabular}{c}
                4.0 (1.6)\\
                4.8 (1.1)
            \end{tabular} 
        \\
        \hline
            \begin{tabular}{c}
                 ``Which is easier to start?''\\
                (1: LED - 7: PUSH)
            \end{tabular} &
            \begin{tabular}{c}
                 8th \\
                 16th 
            \end{tabular} &
            \begin{tabular}{c}
                3.0 \\
                4.5 
            \end{tabular} &
            \begin{tabular}{c}
                3.2 (1.1)\\
                4.5 (1.0)
            \end{tabular} 
        \\
        \hline
            \begin{tabular}{c}
                 ``Which has a higher future potential?''\\
                (1: LED - 7: PUSH)
            \end{tabular} &
            \begin{tabular}{c}
                 8th \\
                 16th 
            \end{tabular} &
            \begin{tabular}{c}
                5.0 \\
                5.5 
            \end{tabular} &
            \begin{tabular}{c}
                4.8 (0.4)\\
                5.5 (0.5)
            \end{tabular} 
        \\
    \end{tabular}
    \caption{Mean and standard deviation of the seven-point Likert scale questionnaire about the impressions of each presented stimulus}~\label{tb:Croch-7_Liker-scale}
\end{table}

 The results of the seven-point Likert scale questionnaire are presented in Table~\ref{tb:Croch-7_Liker-scale}. The mean of ``Ease of recognition'' after the 16th procedure time is larger (better) than that after the 8th procedure time for each stimulus. Furthermore, the ``Ease of recognition'' of the push-type is slightly better than that of the LED-type after the 16th procedure time. The results for ``Which is easier to recognize?'' of the push-type were better than those of the LED-type after the 16th procedure time.
 These results indicate the possibility that all participants became used to the push-type. The result of ``Which is easier to start'' shows that it had been solving the problem of ``Difficult to clearly perceive the start signal.'' Moreover, the results for ``Which has a higher future potential?'' shows that all participants expect the future potential of the push-type to be better than that of the LED-type. These results confirm that the push-type device befits the start signal for DHH sprinters better than the LED-type.

\subsubsection{Comments of Questionnaire and Interview}
 The participant's opinions collected through questionnaires and interviews are summarized in the following points. 
 
 A few negative opinions, in terms of ``Influence of habituation,'' were received; however, they all changed to positive ones. Although a few participants were worried about push-type devices initially, they became comfortable with using it eventually: ``{\it Was worried about push-type devices at first, but now I don't care}'' (P2), ``{\it now that I have repeated the test a number of times, I feel that the push-type is better}'' (P3), and ``{\it The push-type device made it easier to start than the LED after using the device a number of times}'' (P5).
 
 A responses provided negative opinions about setting the position of the hand: ``Takes time to set the position of the hand.'' A few participants felt that ``{\it the push-type devices cannot be fine-tuned once fixed or it takes time to adjust the position because it slightly shifts if done once}'' (P4), ``{\it it feels uncomfortable when fixing my hand, and it is better to put your hands naturally with less practice}'' (P5). 
 
 Furthermore, there were also a few positive opinions in terms of the device making it ``easy to concentrate.'' A few participants responded that ``{\it they want to start by feeling instead of looking at the LED (Don't want to be tied to looking at the LED with my eyes), and they feel that it is easy to concentrate because they do not need to look at the LED}'' (P2), and ``{\it I started after being stimulated, and therefore, it was easy to do passively. Furthermore, it became easier to concentrate on the start}'' (P3).
 In terms of ``easier to recognize,'' both positive and negative opinions were received. A few participants felt that ``{\it the haptic device is easy to start because it feels like ``I hurt?''; on the other hand, with an LED, the body does not move as if it were listening to an instruction}'' (P1), ``{\it the start to hear with sound and the start to feel with Push are similar}'' (P2),  and ``{\it the LED was easy to recognize; however, there was a misalignment and I felt that the start was delayed}'' (P3), ``{\it Since it is a judgment by the brain that it is LED, it takes time and effort; since it reacts with the body by Push, I could understand immediately}'' (P5), ``{\it LED is still easier to recognize; I feel a sense of incongruity because I am conscious of my hands. Both are easier to recognize, but the Push needs power. I want to feel like a Bang}'' (P6).

\section{Discussion}

\subsection{Are the experiment designs and results valid for DHH people?}
 In these experiments, we recruited male DHH participants who had the necessary experience and were accustomed to a crouch start in a sprint race because we aimed to address the following three questions: the significant difference caused by the habituation of crouching start~\cite{Yau:Visual_Athletes_NonAthletes}, the habituation to the present stimulus~\cite{Christopher:SwimmingStartRT}, and the suppression of the significant difference caused by gender~\cite{Ferguson:RT_SexBody,Margit:SVRT_SexAge}.
 Furthermore, we examined the RT using the crouch start of the employed HaptStarter and LED-type visual stimulus procedure 16 times per participant and determined if there were any effects of habituation in the first and second halves. We administered a questionnaire and conducted an interview after finishing the 8th and 16th procedure time of the reaction time using the crouch start.
 Furthermore, we examined the RT using the crouch start of the employed HaptStarter and LED-type visual stimulus in the 16 procedure times per participant if there were any effects of habituation in the first and second halves. 
 We also considered that the whole-body RT measurement experiment could be susceptible to the practice effect. Therefore, we employed a practice period before measuring each stimulus of one procedure time to confirm the intentions of the participants.

\subsection{Insights and Opportunities for the HaptStarter Prototype Design}
\subsubsection{Rules and Install to this scene?}
 Currently, technologies related to visual, auditory, and haptic presentation methods are constantly being researched and developed among the technologies in the HCI field. However, it may not consider the problems when introducing the technologies to sports fields. For example, the rules about ``RULE 162 The Start''~\cite{IAAF:RulesStart} are as follows:
 
    \begin{quote}
        In races up to and including 400m (including the first leg of 4$\times$200m, the Medley Relay and 4$\times$400m), a crouch start and the use of starting blocks are compulsory. After the ``On your marks'' command, an athlete shall approach the start line, assume a position completely within his allocated lane and behind the start line. An athlete shall not touch either the start line or the ground in front of it with his hands or his feet when on his mark. Both hands and at least one knee shall be in contact with the ground and both feet in contact with the foot plates of the starting blocks. At the ``Set'' command, an athlete shall immediately rise to his final starting position retaining the contact of the hands with the ground and of the feet with the foot plates of the blocks. Once the Starter is satisfied that all athletes are steady in the “Set” position, the gun shall be fired.
    \end{quote}
 
 Following these rules makes it difficult to use the method presented to both hand's finger pads and both foot's soles. However, these rules can be followed if a sensitive haptic modality can be felt in the first joint such as the sensitive finger pad by we could find. The method to receive and present other sensory information with a device can be used if the feedback time (time to convert sensory information into another sensory information) is short. However, the method is problematic, which includes the presentation time of the feedback time despite the short feedback time.
 A reaction is delayed for the included feedback time compared to that when directly presenting to the senses. 
 We thought it possible to ignore the feedback time by directly receiving the start signal; therefore, we focused on directly presenting sensory information in this research because this target is a start signal.
 Nevertheless, our design's interface is still in the prototype stage, and issues remain. Specifically, how the interface should be applied to athletes with artificial arms or wheelchairs has not yet been considered. In athletes with artificial arms, we need to think about whether the contact interface for the artificial arms should be different from our prototype. In athletes with wheelchairs, we need to discuss whether athletes with wheelchairs can be included in the universal design we aiming because the start method in a sprint race to be used is not a crouch start.

\subsubsection{Is the Optimum Interface the Best Interface?}
 We could identify the easily recognized haptic stimulus interface in many haptic stimuli connected with the first joint of the thumb. Moreover, it is clear that the haptic stimulus is better than the visual stimulus for DHH participants when considering the RT when using the crouch start. 
 However, our optimum interface suggests that we identified the optimum from three interface types, and it is not the best interface for the DHH sprinters. This result considers the comments of the questionnaire and interview.
 Specifically, a negative opinions about ``Setting the position of the hand'' and ``Influence of habituation''. Nevertheless, we cannot complete the analysis that relationship between the RT data and the participant's qualitative feedback. Because we need to hear the participant's background that deaf school experience, training, experience using the optical stimulation start system, etc.
 Thus, we need to improve a point ``Setting the position of the hand'', also think can improve to the optimum interface that requires a haptic modality that a body can quickly move after recognized as ``Bang,'' which is the same as the sound of the gun referring as other prior study that investigated the relationship between emotion and haptic~\cite{EmotionalHaptic}.
 In addition, we need to examine whether hearing sprinters is easily recognizable in this haptic modality, as DHH sprinters. 
 In future work, we must pursue these as each step of the best design for the haptic interface.

\subsubsection{Should the Solenoid Start-Up Time be Shorter than  1 ms?}
 The solenoid used in the developed haptic generating device has 8.7 ms in the mean of the start-up time measured by the acceleration sensor. The RT of the haptic stimulus measurement indicates that RT, which includes the start-up time of the solenoid in two experiments. Thus, we concluded that the original RT can be 8.7 ms faster than this measured RT for the haptic. Among them, in the experiment result using the crouch start, the difference is about 23.8 ms when considering the start-up time of the solenoid (8.7 ms) from the mean of haptic RT and the start-up time of the the LED (several $\mu$s) from the mean of the visual RT. The RT indicates that the difference between the visual and haptic stimuli is substantially the same as the number (about 25 ms) of Ifukube's reports~\cite{Ifukube:SensoryProsthesis}.
 However, the rules about ``Fully Automatic Timing and Photo Finish System''~\cite{IAAF:RulesTiming} are as follows:
 
    \begin{quote}
        The System shall be started automatically by the Starter's signal, so that the overall delay between the report from the muzzle or its equivalent visual indication and the start of the timing system is constant and equal to or less than 0.001 second.   
    \end{quote}
 
 Thus, we must limit the solenoid start-up time that is constant and equal to or less than 1 ms. Alternatively, we need to develop a control device that matches the solenoid start-up time to measure the RT from the solenoid start-up time.

\subsection{Universal Design vs. Assistive Technology?}
\subsubsection{Results indicate that the difference between DHH and Hearing is Minimized? / Really become equality in the sprint race?}
 This point has something to do with what ``Solenoid Start-up Time should be Shorter than 1 ms?.'' 
 The RT result will be almost the same as Ifukube's reports when considering the solenoid start-up time~\cite{Ifukube:SensoryProsthesis} and the haptic RT using the crouch start we measured; this suggests that the difference from the ART may become less than 10 ms. 
 Furthermore, the rules for ``Photo Finish''~\cite{IAAF:RulesTiming} are 
 
    \begin{quote}
        For all races up to and including 10,000m, unless the time is an exact 0.01 second, the time shall be converted and recorded to the next longer 0.01 second, e.g. 26:17.533 shall be recorded as 26:17.54.     
    \end{quote}
 
 This result indicates that the difference between hearing sprinters and DHH sprinters is less likely to appear on the recording. However, we cannot confirm this because we have not been able perform a comparative experiment that includes hearing sprinters or an actual sprint race of athletes. In the future, we should perform and confirm these experiments.
 In addition, we need to think about not only a start signal. In the actual running of the practice, we know that the rhythms, grounding sound, and communication using onomatopoeia remain problematic, as indicated in these studies ~\cite{Saito:PerformanceDHH,Palmer:DeafAthlete}. We must consider that this will make a difference in the accumulation of practice.
 
\subsubsection{Assistive Technology?}
 Currently, DHH can use the light start system based on the rule ``This Rule should also be interpreted so that:'' stated in ``RULE 161 Starting Blocks''~\cite{IAAF:RulesLight}.
 
    \begin{quote}
        The use of lights, by deaf or hearing impaired athletes only, at the start of races is allowed and is not considered assistance. It should however be the obligation of the athlete or his team for the financing and supply of such equipment and its compatibility with the start system in use, unless at a particular meeting where there is an appointed technical partner who can provide it.  
    \end{quote}
 
 We assume that a hearing and DHH sprinter run through 100 m in the same time (10`` 80). The hearing sprinter reacts with the sound and starts, so the record is 10`` 80. The DHH sprinter reacts with light and starts; the record is 10`` 83 of the total run through 100 m, i.e., 10``80 plus the visual delay time of 0`` 03~\cite{Ifukube:SensoryProsthesis}. This case does not realize equality in sports; however, it creates discrimination on record.
 The problem that DHH sprinters have difficulty perceiving audio information during performance is something we should believe about how to solve this problem after considering whether there is any discrimination on the record. Furthermore, the DHH using only light-type visual stimulus fits the ``Assistive Technology'' definition defined in The Assistive Technology Act of 1998 in the United States~\cite{USA:AT}. 
 The suitable term for the concept we aim than this term is ``Universal Design,'' which was defined by the proponent Ronald L. Mace~\cite{CenterForUD:About}  as 
 
    \begin{quote}
        Universal design is the design of products and environments to be usable by all people, to the greatest extent possible, without the need for adaptation or specialized design.
        -- Ron Mace
    \end{quote}
 
 This definition continues to be used today~\cite{UN:UD, CEUD:UD, GSA:UD}. 
 The Ronald L. Mace announcement at ``Designing for the 21st Century: An International Conference on Universal Design''~\cite{CenterForUD:LastSpeech} also touched on the differences between ``Universal Design'' and ``Assistive Technology'' as
  
    \begin{quote}
        Now, assistive technology to me is really personal use devices—those things focused on the individual—things that compensate or help one function with a disability.
    \end{quote}
 
 Thus, if we advocate equality in sports, the DHH athletes using only light-type visual stimulus sways towards the ``Assistive Technology'' concept, which should be avoided as far as possible. We need to believe introducing a system that everyone can use easily (``Universal Design’’). Therefore, if a light-type visual stimulus is used as ``Universal Design,’’ it will not be suitable for the blind and low-vision people. The reason for this is violating ``PRINCIPLE ONE: Equitable Use,''  ``PRINCIPLE THREE: Simple and Intuitive Use,'' and ``PRINCIPLE FOUR: Perceptible Information'' of the ``Universal Design'' principles~\cite{CenterForUD:Principles}. However, if it is a haptic stimulus, it has the potential to follow the ``Universal Design'' principles. 
 We position the HaptStarter as a start signal that re-standardizes the definition of ``Universal Design'' instead of ``Assistive Technology'' because we believe in advocating for sports equality.
  
    \begin{figure}[t]
        \centering
        \includegraphics[width=1.0\columnwidth]{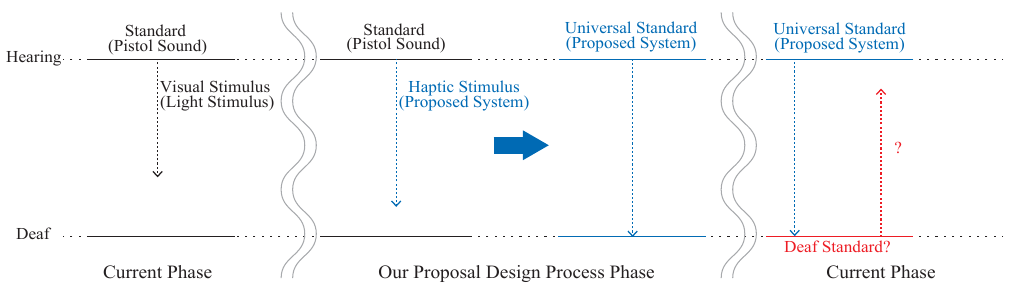}
        \caption{Phase proposal and universal standard; what is the deaf Standard?}
        \label{fig:UniversalStandard}
    \end{figure} 

 However, we also need to simultaneously think about where it is necessary to consider the base point of the universal standard, which plays the role of universal design. Assuming that the ``Current phase'' will change to ``Our Proposal Design Process Phase,'' as indicated in Figure~\ref{fig:UniversalStandard}, this universal standard will only start from hearing sports. It will eventually fully expand to meet the universal design principles. 
 In such a case, can the significance of deaf sports simply be a minor change from hearing sports to DHH persons? Furthermore, should we consider presentation using visual and haptic stimuli as deaf standards of a new deaf sports?

\section{Conclusion and Future works}
 We developed the contact interface of the push-type haptic stimulus befit a brand-new haptic stimulus start system befit DHH sprinters, and we adapted it for a crouch start. In the experiment measuring the RT using the button press, we identified the method to use the push-type device (2 mm) to contact the first joint of thumb as the optimum contact interface of the push-type haptic stimulus for DHH sprinters. Furthermore, the experiment results of the RT when using the crouch start confirmed that the optimum contact interface of the push-type haptic stimulus was valid when compared to the LED-type visual stimulus for the DHH sprinters. 
 The result of the RTs are limited to DHH male people who have experience and are accustomed to the crouch start in a sprint race. Therefore, these results may not necessarily be relevant to daily life; the results may also depend on whether the subjects are gender and healthy.
 In this study, we did not collect the participants' attributes information to verify whether they are healthy.
 We cannot ensure that we can be used in future studies about simple and whole-body RTs because the RTs measured in this study are different from the previous study~\cite{IRWIN:SRT_Standard}.
 We focused on the finger pad and the first joint for haptic stimuli and we compared a simple RT of each. Furthermore, the whole-body RT measurement system is different from the system used in the athletics sprint race because we developed the system.
 Therefore, to continue this research project, we need to unify the system for measuring the simple RT and to use the system in athletics sprint races when measuring the whole body RT.
 Nevertheless, as shown in other prior studies~ \cite{Petermeijer:DriverRT, Gray:HapticRT_Drive, Itoh:HapticRT_Drive, Jordan1:HapticRT_Drive, Jordan2:HapticRT_Drive, Mohebbi:HapticRT_Drive, Straughn:HapticRT_Drive, Scott:HapticRT_Drive}, an RT is fast by a haptic support system. This result can also be applied to the athletics sprint race's start signal.
 Thus, we believe that we will have positioned this device to have a trigger for re-standardization with the concept of universal and not only assistive technology.
 In the future, we will first improve the HaptStarter to update the interface that can more clearly detect the reaction for the DHH and hearing sprinters based on the ``PRINCIPLE THREE: Simple and Intuitive Use” and ``PRINCIPLE FOUR: Perceptible Information” of the ``Universal Design” principles, and the first step is increasing the number of DHH male and female participants in the RT measurement experiment. The second step is conducting the RT measurement experiment for hearing male and female participants. In addition, we verify whether there is a significant difference in the RT between the hearing and DHH top sprinters by measuring the RT for the hearing and DHH top sprinters in a thorough experiment environment using a force plate. 
 Second, we plan to verify that the DHH, the blind and low-vision, and the hearing people can all equally perceive a reaction without noise according to the ``PRINCIPLE ONE: Equitable Use” of the ``Universal Design” principles. These approaches are the next step of creating innovations that can clearly detect the start signal, which the DHH, the blind and low-vision, and the hearing people can all equally perceive reaction without noise. In other words, we will attempt to re-standardize the haptic stimuli based on the ``Universal Design" principle for everyone from an auditory stimulus in the start signal and referee's signal in competitive sports. Although parallel, we need to think about how to change the rules while exchanging information with World Athletics and other organizations.

\section{Acknowledgements}
 This work was supported by JST CREST Grant Number JPMJCR19F2, Japan and Promotional Projects for Advanced Education and Research in the National University Corporation of Tsukuba University of Technology (NTUT).

\bibliography{mybibfile}

\end{document}